\documentclass[a4paper,12pt]{article}
\usepackage {amsmath}
\usepackage {amsxtra}
\usepackage {amsbsy}
\usepackage {amscd}
\usepackage {latexsym}
\usepackage {amssymb}
\usepackage {amsfonts}
\usepackage {a4wide}
\usepackage {indentfirst}
\usepackage {bm}
\usepackage{graphicx}

\newcommand {\oks}[2]{{\raise0.7ex\hbox{${\scriptstyle #1}$}\!\mathord{\left/
{\vphantom{{1}{2}}}\right.\kern-\nulldelimiterspace}\!\lower0.7ex
\hbox{${\scriptstyle #2}$}}}

\large

\begin{document}

\title{Neutrino quantum states and spin light in matter}
\author{Alexander Studenikin{\thanks{
Address: Department of Theoretical Physics, Moscow State
University, 119992  Moscow, Russia,
 e-mail:studenik@srd.sinp.msu.ru .}},
Alexei Ternov{\thanks{Department of Theoretical Physics, Moscow
Institute for Physics and Technology, 141700 Dolgoprudny, Russia,
e-mail:A\_Ternov@mail.ru }}}
  \date{}
  \maketitle

\begin{abstract}{On the basis of the exact solutions of the
modified Dirac equation for a massive neutrino moving in matter we
develop the quantum theory of the spin light of neutrino
($SL\nu$). The expression for the emitted photon energy is derived
as a function of the density of matter for different matter
compositions. The dependence of the photon energy on the
helicities of the initial and final neutrino states is shown
explicitly. The rate and radiation power of the $SL\nu$ in matter
are obtained with the emitted photon linear and circular
polarizations being accounted for. The developed quantum approach
to the $SL\nu$ in matter (which is similar to the Furry
representation of electrodynamics) can be used in the studies of
other processes with neutrinos in the presence of matter.}

\end{abstract}

\section {Introduction}

The problem of a neutrino propagation through the background
matter has attracted the permanent interest for many years. The
crucial importance of the matter effects was demonstrated in the
studies of Refs.\cite{WolPRD78,MikSmiYF85} where the resonance
amplification of the neutrino flavour oscillations in the presence
of the matter (the Mikheyev-Smirnov-Wolfenstein effect) was
discovered. The similar resonance effect in the neutrino spin
oscillations in matter was considered for the first time in
\cite{LimMarPRD88,AkhPLB88}.

It is a common knowledge that the matter effects in neutrino
oscillations have a large impact on solar-neutrino problem (see
for a recent review  \cite{BahGonPenhp0406294}); it may be of
interest in the context of neutrino oscillation processes in
supernovae and neutron stars \cite{FulMayWilSchAPJ87}. It is also
believed that the neutrino oscillations in the presence of matter
play important role in the early Universe \cite{DolYF81}. The
recent global analysis of the neutrino oscillation data can be
found in \cite{MalValSchTorVal04}.

The neutrino interaction with matter can bring about new phenomena
that do not exist in the vacuum. In particular, we have recently
shown \cite{LobStuPLB03,LobStuPLB04} that a massive neutrino
moving in the background matter and electromagnetic fields can
produce a new type of the electromagnetic radiation. We have named
this radiation as the "spin light of neutrino" ($SL\nu$)
\cite{LobStuPLB03} in order to manifest the correspondence with
the magnetic moment dependent term in the radiation of an electron
moving in a magnetic field (see \cite{TerUFN95}). The
electromagnetic radiation of a neutrino moving in a magnetic field
(the $SL\nu$ in a magnetic field) was studied before in
\cite{BorZhuTer88}. Recently we have also considered
\cite{DvoGriStuIJMP04} the $SL\nu$ in gravitational fields.

In \cite{LobStuPLB03,LobStuPLB04} we develop the quasi-classical
theory of the $SL\nu$ on the basis of the quasi-classical
description \cite{EgoLobStu00LobStu01DvoStuJHEP02StuPAN04} of the
neutrino spin evolution in the background matter. However, the $
SL\nu$, as it is explicitly shown below, is a quantum process
which originates from the quantum spin flip transitions from the
"exited"  helicity  state to  the  low-lying  helicity neutrino
state in matter. It is very important to develop the quantum
theory of this phenomena because the quasi-classical approach
gives only an approximate description.   In  particular, within
the quasi-classical approach it is not possible  to derive one  of
the main quantities such  as the  emitted photon  energy (this
quantity was obtained in  our earlier paper \cite{LobStuPLB03}
from an estimation  based  on the dimensional reasoning). The
quantum treatment of the $SL\nu$ in matter developed below (see
also \cite{StuTerhep_ph_0410296_97}) enables us to get the
expression for the emitted photon energy which explicitly shows
the dependence on the initial and final neutrino helicities and
which also accounts for the dependence on the direction of the
radiation. It should be noted that the photon energy is
proportional to the density of matter, it exhibits significant
dependence on the type of the neutrino and also on the background
matter composition.

In the performed below quantum calculations of the $SL\nu$
radiation rate and power we also account for the emitted photon's
(linear and circular) polarizations. The polarization properties
of the radiation can be important for experimental observations of
the $SL\nu$.

 More
generally, one  of  the goals of the paper is to show how the well
established Furry  representation \cite{FurPR51}, previously used
for description of quantum    processes   in   the   presence   of
external electromagnetic  fields,  can  be  used  in  the  case of
processes  with neutrinos in the presence  of  matter. To reach
this goal, the solution of the modified Dirac equation for a
neutrino in the presence of matter is derived (see also
\cite{StuTerhep_ph_0410296_97}) in explicit and simple form
appropriate for the further use in quantum calculations (within
the representation which is similar to the Furry representation)
of different characteristics of quantum processes with
participation of neutrinos in matter.

\section{Dirac equation for neutrino in matter}
To derive the quantum equation for the neutrino wave function in
the background matter we start with the effective Lagrangian that
describes the neutrino interaction with particles of the
background matter. For definiteness, we consider the case of the
electron neutrino $\nu$ propagating through moving and polarized
matter composed of only electrons (the electron gas).  Assume that
the neutrino interactions are described by the extended standard
model supplied with $SU(2)$-singlet right-handed neutrino
$\nu_{R}$. We also suppose that there is a macroscopic amount of
electrons in the scale of a neutrino de Broglie wave length.
Therefore, the interaction of a neutrino with the matter
(electrons) is coherent. In this case the averaged over the matter
electrons addition to the vacuum neutrino Lagrangian, accounting
for the charged and neutral interactions, can be written in the
form
\begin{equation}\label{Lag_f}
\Delta L_{eff}=-f^\mu \Big(\bar \nu \gamma_\mu {1+\gamma^5 \over
2} \nu \Big), \ \  f^\mu={G_F \over \sqrt2}\Big((1+4\sin^2 \theta
_W) j^\mu - \lambda ^\mu \Big),
\end{equation}
where the electrons current $j^{\mu}$ and electrons polarization
$\lambda^{\mu}$ are given by
\begin{equation}
j^\mu=(n,n{\bf v}), \label{j}
\end{equation}
and
\begin{equation} \label{lambda}
\lambda^{\mu} =\Bigg(n ({\bm \zeta} {\bf v} ), n {\bm \zeta}
\sqrt{1-v^2}+ {{n {\bf v} ({\bm \zeta} {\bf v} )} \over
{1+\sqrt{1- v^2}}}\Bigg),
\end{equation}
$\theta _{W}$ is the Weinberg angle.

 The Lagrangian (\ref{Lag_f})
accounts for the possible effect of the matter motion and
polarization. Here $n$, ${\bf v}$, and ${\bm \zeta} \ (0\leqslant
|{\bm \zeta} |^2 \leqslant 1)$ denote, respectively, the number
density of the background electrons, the speed of the reference
frame in which the mean momenta of the electrons is zero, and the
mean value of the polarization vector of the background electrons
in the above mentioned reference frame. The detailed discussion on
the determination of the electrons polarization can be found in
\cite{EgoLobStu00LobStu01DvoStuJHEP02StuPAN04}.

From the standard model Lagrangian with the extra term $\Delta
L_{eff}$ being added, we derive \cite{StuTerhep_ph_0410296_97} the
following modified Dirac equation for the neutrino moving in the
background matter,
\begin{equation}\label{new} \Big\{
i\gamma_{\mu}\partial^{\mu}-\frac{1}{2}
\gamma_{\mu}(1+\gamma_{5})f^{\mu}-m \Big\}\Psi(x)=0.
\end{equation}
This is the most general equation of motion of a neutrino in which
the effective potential $V_{\mu}=\frac{1}{2}(1+\gamma_{5})f_{\mu}$
accounts for both the charged and neutral-current interactions
with the background matter and also for the possible effects of
the matter motion and polarization. It should be noted here that
the modified effective Dirac equations for a neutrino with various
types of interactions with the  background environment  were used
previously in \cite{ ChaZiaPRD88, ManPRD88,NotRafNPB88, NiePRD89,
HaxZhaPRD91,WeiKiePRD97} for the study of the neutrino dispersion
relations and derivation of the neutrino oscillation probabilities
in matter. If we neglect the contribution of the neutral-current
interaction and possible effects of motion and polarization of the
matter then from (\ref{new}) we can get corresponding equations
for the left-handed and right-handed chiral components of the
neutrino field derived in \cite{PanPLB91-PRD92}.

The generalizations of the modified Dirac equation (\ref{new}) for
more complicated matter compositions (and the other flavour
neutrinos) are just straightforward. For instance, if one
considers the case of realistic matter composed of electrons,
protons and neutrons, then the matter term in (\ref{new}) is (see,
for instance, \cite {PalIJMPA92} and the fourth paper of
ref.\cite{EgoLobStu00LobStu01DvoStuJHEP02StuPAN04})
\begin{equation}\label{f_mu}
f^\mu={G_F \over \sqrt2}\sum\limits_{f=e,p,n}
j^{\mu}_{f}q^{(1)}_{f}+\lambda^{\mu}_{f}q^{(2)}_{f}
\end{equation}
where
\begin{equation}\label{q_f}
  q^{(1)}_{f}=
(I_{3L}^{(f)}-2Q^{(f)}\sin^{2}\theta_{W}+\delta_{ef}), \ \
q^{(2)}_{f}=-(I_{3L}^{(f)}+\delta_{ef}), \ \ \delta_{ef}=\left\{
\begin{tabular}{l l}
1 & for {\it f=e}, \\
0 & for {\it f=n, p}, \\
\end{tabular}
\right.
\end{equation}
and $I_{3L}^{(f)}$ is the value of the isospin third component of
a fermion $f$, and $Q^{(f)}$ is the value of its electric charge.
The fermion's currents $j^{\mu}_{f}$ and polarizations
$\lambda^{\mu}_{f}$ are given by eqs.(\ref{j}) and (\ref{lambda})
with the appropriate substitutions: $n, \bm \zeta, {\bf v}
\rightarrow n_f, {\bm \zeta}_f, {\bf v}_f$.

\section{Neutrino wave function and energy spectrum in matter}
In the further discussion below we consider the case when no
electromagnetic field is present in the background. We also
suppose that the matter is unpolarized, $\lambda^{\mu}=0$.
Therefore, the term describing the neutrino interaction with the
matter is given by
\begin{equation}\label{f}
f^\mu=\frac{\tilde{G}_{F}}{\sqrt2}(n,n{\bf v}),
\end{equation}
where we use the notation $\tilde{G}_{F}={G}_{F}(1+4\sin^2 \theta
_W)$.

In the rest frame of the matter the equation (\ref{new}) can be
written in the Hamiltonian form,
\begin{equation}\label{H_matter}
i\frac{\partial}{\partial t}\Psi({\bf r},t)=\hat H_{matt}\Psi({\bf
r},t),
\end{equation}
where
\begin{equation}\label{H_G}
  \hat H_{matt}=\hat {\bm{\alpha}} {\bf p} + \hat {\beta}m +
  \hat V_{matt},
\end{equation}
and
\begin{equation}\label{V_matt}
\hat V_{matt}= \frac{1}{2\sqrt{2}}(1+\gamma_{5}){\tilde {G}}_{F}n,
\end{equation}
here $\bf p$ is the neutrino momentum. We use the Pauli-Dirac
representation for the Dirac matrices $\hat {\bm \alpha}$ and
$\hat {\beta}$, in which
\begin{equation}\label{a_b}
    \hat {\bm \alpha}=
\begin{pmatrix}{0}&{\hat {\bm \sigma}} \\
\hat {{\bm \sigma}}& {0}
\end{pmatrix}=\gamma_0{\bm \gamma}, \ \ \
\hat {\beta}=\begin{pmatrix}{1}&{0} \\
{0}& {-1}
\end{pmatrix}=\gamma_0,
\end{equation}
where ${\hat { \bm\sigma}}=({ \sigma}_{1},{ \sigma}_{2},{
\sigma}_{3})$ are the Pauli matrixes.

The form of the Hamiltonian (\ref{H_G}) implies that the operators
of the momentum, $\hat {\bf p}$, and longitudinal polarization,
${\hat{\bf \Sigma}} {\bf p}/p$, are the integrals of motion. So
that, in particular, we have
\begin{equation}\label{helicity}
  \frac{{\hat{\bf \Sigma}}{\bf p}}{p}
  \Psi({\bf r},t)=s\Psi({\bf r},t),
 \ \ {\hat {\bm \Sigma}}=
\begin{pmatrix}{\hat {\bm \sigma}}&{0} \\
{0}&{\hat {\bm \sigma}}
\end{pmatrix},
\end{equation}
where the values $s=\pm 1$ specify the two neutrino helicity
states, $\nu_{+}$ and  $\nu_{-}$. In the relativistic limit the
negative-helicity neutrino state is dominated by the left-handed
chiral state ($\nu_{-}\approx \nu_{L}$), whereas the
positive-helicity state is dominated by the right-handed chiral
state ($\nu_{+}\approx \nu_{R}$).

For the stationary states of the equation (\ref{new}) we get
\begin{equation}\label{stat_states}
\Psi({\bf r},t)=e^{-i(  E_{\varepsilon}t-{\bf p}{\bf r})}u({\bf
p},E_{\varepsilon}),
\end{equation}
where $u({\bf p},E_{\varepsilon})$ is independent on the
coordinates and time. Upon the condition that the equation
(\ref{new}) has a non-trivial solution, we arrive to the energy
spectrum of a neutrino moving in the background matter:
\begin{equation}\label{Energy}
  E_{\varepsilon}=\varepsilon{\sqrt{{\bf p}^{2}\Big(1-s\alpha \frac{m}{p}\Big)^{2}
  +m^2} +\alpha m} ,
\end{equation}
where we use the notation
\begin{equation}\label{alpha}
  \alpha=\frac{1}{2\sqrt{2}}{\tilde G}_{F}\frac{n}{m}.
\end{equation}
The quantity $\varepsilon=\pm 1$ splits the solutions into the two
branches that in the limit of the vanishing matter density,
$\alpha\rightarrow 0$, reproduce the positive and
negative-frequency solutions, respectively. It is also important
to note that the neutrino energy in the background matter depends
on the state of the neutrino longitudinal polarization, i.e. in
the relativistic case the left-handed and right-handed neutrinos
with equal momenta have different energies.

The procedure, similar to one used for derivation of the solution
of the Dirac equation in vacuum, can be adopted for the case of a
neutrino moving in matter. We apply this procedure to the equation
(\ref{new}) and arrive to the final form of the wave function of a
neutrino moving in the background matter
\cite{StuTerhep_ph_0410296_97}:
\begin{equation}\label{wave_function}
\Psi_{\varepsilon, {\bf p},s}({\bf r},t)=\frac{e^{-i(
E_{\varepsilon}t-{\bf p}{\bf r})}}{2L^{\frac{3}{2}}}
\begin{pmatrix}{\sqrt{1+ \frac{m}{ E_{\varepsilon}-\alpha m}}}
\ \sqrt{1+s\frac{p_{3}}{p}}
\\
{s \sqrt{1+ \frac{m}{ E_{\varepsilon}-\alpha m}}} \
\sqrt{1-s\frac{p_{3}}{p}}\ \ e^{i\delta}
\\
{  s\varepsilon\sqrt{1- \frac{m}{ E_{\varepsilon}-\alpha m}}} \
\sqrt{1+s\frac{p_{3}}{p}}
\\
{\varepsilon\sqrt{1- \frac{m}{ E_{\varepsilon}-\alpha m}}} \ \
\sqrt{1-s\frac{p_{3}}{p}}\ e^{i\delta}
\end{pmatrix} ,
\end{equation}
where the energy $E_{\varepsilon}$ is given by (\ref{Energy}), and
$L$ is the normalization length. In the limit of vanishing density
of matter, when $\alpha\rightarrow 0$, the wave function
(\ref{wave_function}) transforms to the vacuum solution of the
Dirac equation.

The quantum equation (\ref{new}) for a neutrino in the background
matter with the obtained exact solution (\ref{wave_function}) and
energy spectrum (\ref{Energy}) establish a basis for a very
effective method (similar to the Furry representation of quantum
electrodynamics) in investigations of different phenomena that can
appear when neutrinos are moving in the media.

Let us now consider in some detail the properties of a neutrino
energy spectrum (\ref{Energy}) in the background matter that are
very important for understanding of the mechanism of the neutrino
spin light phenomena. For the fixed magnitude of the neutrino
momentum $p$ there are the two values for the "positive sign"
($\varepsilon =+1$) energies
\begin{equation}\label{Energy_nu}
  E^{s=+1}={\sqrt{{\bf p}^{2}\Big(1-\alpha \frac{m}{p}\Big)^{2}
  +m^2} +\alpha m}, \ \ \
 E^{s=-1}={\sqrt{{\bf p}^{2}\Big(1+\alpha \frac{m}{p}\Big)^{2}
  +m^2} +\alpha m},
\end{equation}
that determine the positive- and negative-helicity eigenstates,
respectively. The energies in Eq.(\ref{Energy_nu}) correspond to
the particle (neutrino) solutions in the background matter. The
two other values for the energy, corresponding to the negative
sign $\varepsilon =-1$, are for the antiparticle solutions. As
usual, by changing the sign of the energy, we obtain the values
\begin{equation}\label{Energy_anti_nu}
  {\tilde E}^{s=+1}={\sqrt{{\bf p}^{2}
  \Big(1-\alpha \frac{m}{p}\Big)^{2}
  +m^2} -\alpha m}, \ \ \
  {\tilde E}^{s=-1}={\sqrt{{\bf p}^{2}
  \Big(1+\alpha \frac{m}{p}\Big)^{2}
  +m^2} -\alpha m},
\end{equation}
that correspond to the positive- and negative-helicity
antineutrino states in the matter. The expressions in
Eqs.(\ref{Energy_nu}) and (\ref{Energy_anti_nu}) would reproduce
the neutrino dispersion relations of \cite{PanPLB91-PRD92} (see
also \cite{WeiKiePRD97}), if the contribution of the
neutral-current interaction to the neutrino potential were left
out.

Note that on the basis of the obtained energy spectrum
(\ref{Energy}) the neutrino trapping and reflection, the
neutrino-antineutrino pair annihilation and creation in a medium
can be studied
\cite{ChaZiaPRD88,LoePRL90,PanPLB91-PRD92,WeiKiePRD97}.

\section {Quantum theory of spin light of neutrino in matter}
In this section we should like to use the obtained solutions
(\ref{wave_function}) of the equation (\ref{new}) for a neutrino
moving in the background matter for the study of the spin light of
neutrino ($SL\nu$) in the matter. We develop below the {\it
quantum} theory of this effect. Within the quantum approach, the
corresponding Feynman diagram of the $SL\nu$ in the matter is the
standard one-photon emission diagram with the initial and final
neutrino states described by the "broad lines" that account for
the neutrino interaction with the matter. It follows from the
usual neutrino magnetic moment interaction with the quantized
photon field, that the amplitude of the transition from the
neutrino initial state $\psi_{i}$ to the final state $\psi_{f}$,
accompanied by the emission of a photon with a momentum
$k^{\mu}=(\omega,{\bf k})$ and  polarization ${\bf e}^{*}$, can be
written in the form
\begin{equation}\label{amplitude}
  S_{f i}=-\mu \sqrt{4\pi}\int d^{4} x {\bar \psi}_{f}(x)
  ({\hat {\bm \Gamma}}{\bf e}^{*})\frac{e^{ikx}}{\sqrt{2\omega L^{3}}}
   \psi_{i}(x),
\end{equation}
where $\mu$ is the neutrino magnetic moment, $\psi_{i}$ and
$\psi_{f}$ are the corresponding exact solutions of the equation
(\ref{new}) given by (\ref{wave_function}), and
\begin{equation}\label{Gamma}
  \hat {\bm \Gamma}=i\omega\big\{\big[{\bm \Sigma} \times
  {\bm \varkappa}\big]+i\gamma^{5}{\bm \Sigma}\big\}.
\end{equation}
Here ${\bm \varkappa}={\bf k}/{\omega}$ is the unit vector
pointing the direction of the emitted photon propagation. Note the
the develop approach to the considered process in the presence of
matter is similar to the Furry representation in studies of
quantum processes in external electromagnetic fields.

The integration in (\ref{amplitude}) with respect to time yields
\begin{equation}\label{amplitude}
  S_{f i}=-\mu {\sqrt {\frac {2\pi}{\omega L^{3}}}}
  2\pi\delta(E_{f}-E_{i}+\omega)
  \int d^{3} x {\bar \psi}_{f}({\bf r})({\hat {\bf \Gamma}}{\bf e}^{*})
  e^{i{\bf k}{\bf r}}
   \psi_{i}({\bf r}),
\end{equation}
where the delta-function stands for the energy conservation.
Performing the integrations over the spatial co-ordinates, we can
recover the delta-functions for the three components of the
momentum. Finally, we get the law of the energy-momentum
conservation for the considered process,
\begin{equation}\label{e_m_con}
    E_{i}=E_{f}+\omega, \ \ \
    {\bf p}_{i}={\bf p}_{f}+{\bm \varkappa}.
\end{equation}
Let us suppose that the weak interaction of the neutrino with the
electrons of the background is indeed weak. In this case, we can
expand the energy (\ref{Energy}) over ${\tilde {G}}_{F}n/p\ll 1$
and in the liner approximation get
\begin{equation}\label{Energy_2}
  E\approx E_{0}-sm\alpha \frac{p}{E_{0}}+\alpha m,
\end{equation}
where $E_0=\sqrt{p^2 +m^2}$. Then from the law of the energy
conservation (\ref{e_m_con}) we get for the energy of the emitted
photon
\begin{equation}\label{omega}
  \omega=E_{{i}_{0}}-{E}_{{f}_{0}}+\Delta,
\ \   \ \Delta=\alpha m \frac{p}{E_0}(s_{f}-s_{i}),
\end{equation}
where the indexes $i$ and $f$ label the corresponding quantities
for the neutrino in the initial and final states. From
Eq.(\ref{omega}) and the law of the momentum conservation, in the
linear approximation over $n$, we obtain
\cite{StuTerhep_ph_0410296_97}
\begin{equation}\label{omega_1}
    \omega=(s_{f}-s_{i})\alpha m \frac {\beta}{1-\beta \cos
    \theta},
\end{equation}
where $\theta$ is the angle between ${\bm \varkappa}$ and the
direction of neutrino speed ${\bm \beta}$.

From the above consideration it follows that the only possibility
for the $SL\nu$ to appear is provided in the case when the
neutrino initial and final states are characterized by $s_{i}=-1$
and $s_{f}=+1$, respectively. Thus,  on the basis of the quantum
treatment of the $SL\nu$ in the matter composed of electrons we
conclude, that in this process the left-handed neutrino is
converted to the right-handed neutrino (see also
\cite{LobStuPLB03}). The emitted photon energy is given by
\begin{equation}\label{omega_12}
    \omega=\frac{1}{\sqrt{2}}{\tilde G}_{F}n \frac {\beta}{1-\beta \cos
    \theta}.
\end{equation}
Note that the photon energy depends on the angle $\theta$ and also
on the value of the neutrino speed $\beta$. In the case of
$\beta\approx 1$ and $\theta\approx 0$ we confirm the estimation
for  the emitted photon energy that was obtained earlier
\cite{LobStuPLB03} within the dimensional reasoning.

If further we consider the case of the neutrino moving along the
OZ-axes, we can rewrite the solution (\ref{wave_function}) for the
neutrino states with $s=-1$ and $s=+1$ in the following forms
\begin{equation}\label{wave_function_min}
\Psi_{{\bf p},s=-1}({\bf r},t)=\frac{e^{-i(Et-{\bf p}{\bf
r})}}{\sqrt{2}L^{\frac{3}{2}}}
\begin{pmatrix}{0}
\\
{- \sqrt{1+ \frac{m}{E-\alpha m}}} \
\\
{ 0}
\\
{\sqrt{1- \frac{m}{E-\alpha m}}}
\end{pmatrix} ,
\end{equation}
and
\begin{equation}\label{wave_function_min}
\Psi_{{\bf p},s=+1}({\bf r},t)=\frac{e^{-i(Et-{\bf p}{\bf
r})}}{\sqrt{2}L^{\frac{3}{2}}}
\begin{pmatrix}{ \sqrt{1+ \frac{m}{E-\alpha m}}}
\\
{0} \
\\
{ \sqrt{1- \frac{m}{E-\alpha m}}}
\\
{0}
\end{pmatrix}.
\end{equation}
We now put these wave functions into Eq.(\ref{amplitude}) and
calculate the spin light transition rate in the linear
approximation of the expansion over the parameter ${\tilde
{G}}_{F}n/p$. Finally, for the rate we get
\begin{equation}\label{rate}
  \Gamma_{SL}=\frac{1}{2\sqrt{2}}{\tilde G}_{F}^{3}\mu^{2}n^{3}
  \beta^{3}
\int
  \limits_{}^{}\frac{S\sin \theta}{(1-\beta \cos \theta)^{4}}
   d\theta.
\end{equation}
where
\begin{equation}\label{S}
S=(\cos \theta - \beta)^{2}+
  (1-\beta \cos \theta)^{2}.
\end{equation}
Performing the integrations in Eq.(\ref{rate}) over the angle
$\theta$, we obtain for the rate
\begin{equation}\label{rate_1}
  \Gamma_{SL}=\frac{2\sqrt{2}}{3}\mu^{2}{{\tilde G}_{F}}^{3}
  n^{3}\beta^{3}\gamma^{2}.
\end{equation}
This result exceeds the value of the neutrino spin light rate
derived in \cite{LobStuPLB03} by a factor of two because here the
neutrinos in the initial state  are totally left-hand polarized,
whereas the case of the unpolarized neutrinos in the initial state
was consider in \cite{LobStuPLB03}.

The corresponding expression for the radiation power is
\begin{equation}\label{power}
  I_{SL}=\frac{1}{4}\mu^{2}{\tilde G}_{F}^{4}n^{4}\beta^{4}
\int\limits_{}^{}\frac{S\sin \theta}{(1-\beta \cos \theta)^{5}}
   d\theta.
\end{equation}
Performing the integration, we get for the total radiation power
\begin{equation}\label{power_1}
  I_{SL}=\frac{2}{3}\mu^{2}{{\tilde G}_{F}}^{4}
  n^{4}\beta^{4}\gamma^{4}.
\end{equation}

In the performed above quantum treatment of the $SL\nu$ in the
background matter we confirm the main properties of this radiation
obtained within the quasi-classical treatment \cite{LobStuPLB03,
LobStuPLB04, DvoGriStuIJMP04}. In particular, as it follows from
(\ref{power}), the $SL\nu$ is strongly beamed along the
propagation of the relativistic neutrino. The total power of the
$SL\nu$ in the matter is increasing with the increase of the
background matter density and the neutrino $\gamma$ factor,
$I_{SL}\sim n^{4}\gamma ^{4}$.

From the obtained within the quantum treatment expression
(\ref{omega_12})  one can get an estimation for the emitted photon
energy
\begin{equation} \omega =2.37\times
10^{-7}\left( \frac{n}{10^{30}cm^{-3}}\right) \left(
\frac{E}{m_{\nu }}\right) ^{2}eV.
\end{equation}
It follows that for the matter densities and neutrino energies,
appropriate for the neutron star environments, the range of the
radiated photons energies may span up to the gamma-rays.

Using Eq.(\ref{amplitude}) we can also derive the $SL\nu$ rate and
total power in matter accounting for the photon polarization (see
also \cite{StuTerhep_ph_0410296_97}). If
 $\bf j$ is the unit vector pointing in the direction of the neutrino
propagation, then we can introduce the to vectors
\begin{equation}\label{e_12}
  {\bf e}_1= \frac{[{\bm \varkappa}\times {\bf j}]}
  {\sqrt{1-({\bm \varkappa}{\bf j})^{2}}}, \ \
  {\bf e}_2= \frac{{\bm \varkappa}({\bm \varkappa}{\bf j})-{\bf j}}
  {\sqrt{1-({\bm \varkappa}{\bf j})^{2}}},
\end{equation}
that specify the two different linear polarizations of the emitted
photon.  For these vectors it is easy to get
\begin{equation}\label{e_12a}
{\bf e}_1=\{\sin \phi, -\cos \phi\}, \ \ {\bf e}_2=\{\cos \phi
\cos \theta, \sin \phi \cos \theta, -\sin \theta \}.
\end{equation}
Note that the vector ${\bf e}_1$ is orthogonal to ${\bf j}$.
Decomposing the neutrino transition amplitude (\ref{amplitude}) in
contributions from the two linearly polarized photons, one can
obtain the power of the process with radiation of the polarized
photons. For the two linear polarizations determined by the
vectors ${\bf e}_1$ and ${\bf e}_2$, we get
\begin{equation}\label{power_pol}
    I_{SL}^{(1),(2)}=
\frac{1}{4}{\tilde G}_{F}^{4}\mu^{2}n^{4}\beta^{4}
  \int\limits_{}^{}\frac{\sin \theta}{(1-\beta \cos \theta)^{5}}
 \begin{pmatrix}{S^{(1)}}
\\
{S^{(2)}} \
\end{pmatrix}
d\theta,
\end{equation}
where
\begin{equation}\label{S_12}
S^{(1)}=(\cos \theta - \beta)^{2}, \ \ S^{(2)}=(1-\beta \cos
\theta)^{2}.
\end{equation}
Finally, performing the integration over the angle $\theta$ we get
the total power of the radiation of the linearly polarized photons
\begin{equation}\label{power_1}
  \begin{pmatrix}{I_{SL}^{(1)}}
  \\
{I_{SL}^{(2)}}
\end{pmatrix}
  =\begin{pmatrix}{\frac{1}{3}}
  \\
{1}
\end{pmatrix}\frac{1}{2}\mu^{2}{{\tilde G}_{F}}^{4}
n^{4}\beta^{4}\gamma^{4}.
\end{equation}

It is also possible to decompose the radiation power for the
circular polarized photons. We introduce the two unit vectors for
description of the photons with the two opposite circular
polarizations,
\begin{equation}\label{circ_pol}
  {\bf e}_{l}=\frac{1}{\sqrt 2}({\bf e}_{1}+i{\bf e}_{2})
\end{equation}
where $l=\pm 1$ corresponds to the right and left photon circular
polarizations, respectively. Then, for the power of the radiation
of the circular-polarized photons we get
\begin{equation}\label{power_pol_circ}
    I_{SL}^{(l)}=\frac{1}{4}\mu^{2}{\tilde G}_{F}^{4}n^{4}\beta^{4}
\int\limits_{}^{}\frac{\sin \theta}{(1-\beta \cos \theta)^{5}}
 {S^{(l)}}
d\theta,
\end{equation}
where
\begin{equation}\label{S_l}
S^{(l)}=\frac{1}{2}(S^{(1)}+S^{(2)})-l\sqrt{S^{(1)}S^{(2)}}.
\end{equation}
Integration over the angle $\theta$ in (\ref{power_pol_circ})
yields
\begin{equation}\label{power_l}
  {I_{SL}^{(l)}}
    =\frac{1}{3}\mu^{2}{{\tilde G}_{F}}^{4}
n^{4}\beta^{4}\gamma^{4}
    \Big(1-\frac{1}{2}l\beta\Big).
\end{equation}

Information on the photons polarization can be important for the
experimental observation of the $SL\nu$ from different
astrophysical and cosmology objects and media.

\section{Conclusions}

 The developed above quantum theory of the neutrino motion
in the background matter also reveals the nature of the $SL\nu$.
In particular, the application of the quantum theory to this
phenomenon enables us to demonstrate that the $SL\nu$ appears due
to the two subdivided phenomena: (i) the shift of the neutrino
energy levels in the presence of the background matter, which is
different for the two opposite neutrino helicity states, (ii) the
radiation of the $SL\nu$ photon in the process of the neutrino
transition from the "exited" helicity state to the low-lying
helicity  state
 in matter. Therefore, the existence of the
neutrino-spin self-polarization effect
\cite{LobStuPLB03,LobStuPLB04} in the matter is confirmed within
the solid base of the developed quantum approach\footnote{The
neutrino-spin self-polarization effect in the magnetic and
gravitational fields were discussed in \cite {BorZhuTer88} and
\cite{DvoGriStuIJMP04}, respectively.}.

From the derived within the quantum approach photon energy
(\ref{omega_1}) and expressions (\ref{f_mu}) and (\ref{q_f}) it is
just straightforward the critical dependence of the $SL\nu$
radiation on the type of the neutrino flavour and the background
matter composition. For instance, in the case of the electron
neutrino and a realistic matter composition, when protons and
neutrons are also present in the background in addition to
electrons, for the value of $\alpha_{\nu_e}$ from (\ref{f_mu}) and
(\ref{q_f}) one gets
\begin{equation}\label{alpha}
  \alpha_{\nu_e}=\frac{1}{2\sqrt{2}}\frac{G_F}{m}\Big(n_e(1+4\sin^2 \theta
_W)+n_p(1-4\sin^2 \theta _W)-n_n\Big),
\end{equation}
where $n_{e,p,n}$ denote, respectively, the number density of
electrons, protons and neutrons. For electrically neutral and
neutron reach matter $\alpha_{\nu_e}<0$. Thus, in this case the
spin-light radiation from the relativistic left-handed electron
neutrino is suppressed. However, due to the fact that for the
anti-neutrino the above value of $\alpha_{\nu_e}$ changes sign,
the electron anti-neutrino can effectively produce the $SL\nu$ in
such a matter.

For the muon and tau neutrinos moving in the same matter, one can
get from (\ref{f_mu}) and (\ref{q_f}) that
\begin{equation}\label{alpha}
  \alpha_{\nu_\mu,\nu_\tau}=
  \frac{1}{2\sqrt{2}}\frac{G_F}{m}\Big(n_e(4\sin^2 \theta
_W-1)+n_p(1-4\sin^2 \theta _W)-n_n\Big).
\end{equation}
Therefore, from the expression (\ref{omega_1}) for the emitted
photon energy we conclude that the relativistic muon and tau
neutrinos can not effectively produce the spin light in the
electrically neutral matter composed of neutrons. However, for the
muon and tau anti-neutrinos $\alpha_{\nu_\mu,\nu_\tau}>0$ in the
considered matter, therefore the spin light can be emitted in this
case. Note that eqs.(\ref{power_pol_circ}-\ref{power_l}) can be
used for evaluation of the radiation power of the circular
polarized photons if the following substitution is made:
$l\rightarrow-l$.

  Finally, the developed quantum
method of accounting for a matter background in processes with
neutrinos, which is similar to the Furry representation for
processes in external electromagnetic fields, would have a large
impact on the studies of different processes with neutrinos
propagating in the astrophysical and cosmology media.

\section{Acknowledgement}
The authors thanks Alexander Grigoriev and Andrey Lobanov for
discussion on some of the issues of the present paper.

\end{document}